**Essential properties of $AlCl_4^-$/ $AlCl_4$ -graphite intercalation compounds of aluminum-ion-based battery cathodes**

Wei-Bang Li[1], Ming-Hsiu Tsai[1], Shih-Yang Lin[2], Kuang-I Lin[3], Ming-Fa Lin[1,4*]

[1]Department of Physics, National Cheng Kung University, Tainan, Taiwan

[2]Department of Physics, National Chung Cheng University, Chiayi, Taiwan

[3]Center for Micro/Nano Science and Technology, National Cheng Kung University, Tainan, Taiwan

[4]Hierarchical Green-Energy Materials (Hi-GEM) Research Center, National Cheng Kung University, Tainan, Taiwan

Email: mflin@ncku.edu.tw

**Abstract**

Up to now, many guest atoms/molecules/ions have been successfully synthesized into graphite to form the various compounds. For example, alkali-atom graphite intercalation compounds are verified to reveal the stage-n structures, including $LiC_{6n}$ and $LiM_{8n}$ [M=K. Rb and Cs; n=1, 2, 3; 4]. On the other side, $AlCl_4^-$-ion/$AlCl_4$-molecule ones are examined to show stage-4 and stage-3 cases at room and lower temperatures, respectively. Stage-1 and stage-2 configurations, with the higher intercalant concentrations, are unable to synthesize in experimental laboratories. This might arise from the fact that it is quite difficult to build the periodical arrangements along the longitudinal $\hat{z}$ and transverse directions simultaneously for the large ions or molecules. Our works are mainly focused on stage-1 and stage-2 systems in terms of geometric and electronic properties. The critical features, being associated with the atom-dominated energy spectra and wave function within the specific energy ranges, the active multi-orbital hybridization in distinct chemical bonds, and atom- & orbital-decomposed van Hove singularities, will be thoroughly clarified by the delicate simulations and analyses.

**Introduction**

A pristine graphite and its intercalation/de-intercalation compounds with guest atoms/molecules/ions [1] display the rich geometric symmetries, mainly owing to the chemical modifications [2]. The Bernal graphite consist of a periodical carbon-honeycomb lattices along the z-direction through the AB stacking, i.e., it is the so-called AB-stacked bulk graphite. Each graphitic sheet remains a planar structure, clearly illustrating the orthogonal features of the significant $\pi$ and $\sigma$ bondings. The former, which is due to the carbon-$2p_z$ orbital hybridizations, could survive in the intralayer [3] and interlayer [van Der Waals;[4]] atomic interactions. It is responsible for the low-energy electronic properties of a 3D graphite [Fig. 2(b); [5]]. However, the strength of the latter remains the same even in the presence of interlayer couplings. As a result, it is easy to distinguish the $\sigma$ and $\pi$ electronic states. In general, this phenomenon keeps unchanged during the chemical reaction processes, i.e., planar graphene layers exist under the very strong $sp^2$-$\sigma$ bondings.

Apparently, the sufficiently wide spacing of ~3.35 Å between two neighboring graphene layers, which are created by the weak, but significant van der Waals interactions [the interlayer $2p_z$ orbital hybridizations; [6], [7]], are available for the easy intercalations of the guest $AlCl_4^-$ ions and $AlCl_4$ molecules intercalations. The interlayer distance [$I_c$=3.35Å under a pristine case; [8], [9]] is greatly enhanced under the various chemical environments, as clearly indicated in Tables 1 and 2 for ion and molecule intercalations, respectively. It is very sensitive the change of intercalation concentration and arrangement [Fig. 1]. As to the ion/molecule cases, $I_c$s are, respectively, 11.30, 11.29, 10.81 and 10.65Å/8.77, 8.78, 8.81 and 8.80 Å for 1:18, 1:24, 1:36 and 1:54 concentrations in terms of ratio about $AlCl_4^-$ / $AlCl_4$ and carbon. A very large $I_c$ clearly indicates the thorough disappearance of the interlayer $2p_z$ orbital couplings. Very interesting, the multi-orbital hybridizations in C- $AlCl_4^-$ or C- $AlCl_4$ bonds can account for the interlayer atomic interactions. As a result, there exist the intralayer C-C bonds, the intra-ion/intra-molecule bondings of $AlCl_4^-$ / $AlCl_4$ , the inter-ion/inter-molecule ones, and the carbon-intercalant interactions . The active orbital hybridizations in distinct chemical bondings need to be identified from the other physical quantities [Figs. 2-4; [10]]. In addition, whether the van der Waals interactions could survive in large-$I_c$ graphite intercalation compounds requires a very detailed numerical examination.

The regular arrangements of large ions/molecules, which possess the high projection symmetries, are chosen for a model study. Very interesting, a pristine graphite has a periodical AB stacking configuration along the z-direction, being dramatically transformed into the AA one during the strong chemical intercalations/de-intercalations. This will lead to the drastic changes of the other fundamental properties [11]. As for each intercalant. aluminum atom is just situated at the hollow site above a hexagon [the top view of the x-y plane projection in Fig 1]; furthermore, it is accompanied two chloride atoms along the dimer & bridge-middle directions simultaneously. The periodical ion/molecule distribution fully occupies the whole interlayer spacing so that the intercalant layer is formed after the chemical modification [12] [13]. This is the so-called stage-1 graphite intercalation compound, where there is only monolayer graphene between two neighboring intercalant layers. And then, the intercalant concentrations decline within the larger unit cells. There are larger Moire superlattices under the dilute cases [14]. These graphite de-intercalation systems are expected to have the lower lattice symmetries and thus more complicated calculations/phenomena [15], compared with alkali atoms [16]. Most important, the atomic configurations in the saturated ions and the unbalanced molecules [the excited angling bonds] [17] play a critical role about an obvious difference in the interlayer distances [~11Å in Table 1 and ~8.8 Å], since the carbon-intercalant interactions are weaker under the former chemical environment. During the concentration variation of large ions/molecules, the sensitive dependences of their internal bonding angles indicate the significant contributions to the total ground state energies. Specifically, the ion-ion interaction might strongly modify the optimal interlayer distances.

The intercalant configurations deserve a closer examination. The previous experimental [18] and theoretical [19] studies have proposed the characterizations of stage-n systems during the chemical modification processes, mainly owing to the unchanged $\sigma$ bonding honeycomb lattices [20] and the drastic changes of free carrier densities [21]. As for n≥2 cases, intercalants exhibit a periodical distribution along the z-direction, but a non-uniform one. However, this unusual configuration might not agree with the natural ion/molecule diffusion phenomena under the external factors [e.g., pressure and thermal energy; 22]. It is mission impossible to create the critical mechanisms in forbidding their intercalations and de-intercalations inside any spacings of the nearest-neighbor graphitic layers. That is, the guest-intercalant transport, which obeys the thermal dynamical laws [23], will be revealed in the successfully synthesized compounds. Obviously, there are certain important differences between the stage-n graphite intercalation compounds and the stage-1 systems [24] [25] with the various concentrations, covering the active chemical bonds, the interlayer distances, the crystal symmetries of Moire superlattices, the bonding angles, the charge transfers, and the metallic or semiconducting behaviors. How to clarify which kind of stacking configuration is the optimal one after the experimental synthesis. This interesting issue could be settled through the method of molecular dynamics [26], in which the physical/chemical/materials environments are the necessary conditions for the delicate numerical simulations. The systematic investigations are required in the near-future basic science researches.

Very apparently, the high-precise X-ray diffraction spectroscopy, as clearly illustrated in Chap.3., is reliable in fully exploring the optimal crystal structures of $AlCl_4^-$/ $AlCl_4$ graphite intercalation compounds. The examined quantities cover the periodical distances along the z-direction and the lattice constants on the (x, y) plane. Whether this method could detect the intercalant-dependent bonding angles is worthy of further thorough investigations. The up-to-date X-ay patterns claim the successful observations about the stage-3 and stage-4 large-intercalant graphite intercalation compounds [18]. The former/the latter is deduced to be relatively stable at lower/room temperatures [~250 K/300 K] during the charging and discharging processes in aluminum-ion-based batteries [18]. Apparently, temperature is one of the critical factors in determining the lattice symmetries. For example, the thermal excitation energies are expected to be comparable with the interlayer graphene-intercalant interactions. The theoretical predictions on the stage-1 systems with the various intercalant concentrations could be generalized to stage-n ones. The stacking configurations would strongly modify the similar physical and chemical phenomena [the great enhancement or reduce of the similar quantities]. This is under a current study [27].

**Theoretical calculations**

The first-principles simulations within DFT by solving the Kohn-Sham equations are dominating methods for the fundamentals properties of periodic systems, i.e., they are frequently utilized to study the geometric, electronic, magnetic, and optical properties. The Perdew-Burke-Ernzerh formula is utilized to deal with many-particle Coulomb effects. The first Brillouin zone is sampled by 9 × 9 × 9 and 100 × 100 × 100 k-point meshes within the Monkhorst-Pack scheme, respectively, for the optimal geometry and band structure. Moreover, the convergence condition of the ground state energy is set to be ~$10^{-5}$ eV between two consecutive evaluation steps, where the maximum Hellmann-Feynman force for each ion is below 0.01 eV/Å during the atom relaxations.

**Results and discussions**

*Rich and unique electronic properties*

Electronic properties of $AlCl_4^-$ -/$AlCl_4$-graphite intercalation compounds, 3D band structures, charge density distributions and density of states, are fully explored by the numerical VASP calculations and further generalized by the phenomenological models. The critical mechanisms, the intralayer $\pi$ & $\sigma$ bondings, the interlayer van Der Waals interactions, the carbon-intercalant orbital hybridizations, and the intra- & inter-ion/inter-molecule interactions, are examined and identified from the delicate analyses. These will be revealed in the semi-metallic or metallic behaviors [the density of free carriers due to the weak or strong charge transfer], the well-characterized/unusual $\pi$ & $\sigma$ bands, the C-, Al- and Cl-dominated energy spectra at the different energy range, the spatial orbital bondings after and before intercalations, and the merged special structures of the energy-dependent van Hove singularities.

Bernal graphite and $AlCl_4^-$ , $AlCl_4$ graphite intercalation compounds, as clearly shown in Fig. 2, present the diverse electronic energy spectra and wave functions. Their first Brilloun zone in Fig 2(a) possesses a hexagonal symmetry [28], in which the (kx, ky)-projection is similar to that of a layered graphene [29]. Band structure of a pristine graphitic system [Fig. 2(b)], being illustrated along the high-symmetry points [ΓKMΓAHLA; 30], is rich and unique. This system is an unusual semimetal, while a monolayer graphene is a zero-gap semiconductor with a vanishing density of states at the Fermi level [31]. Such difference obviously indicates the important role of the interlayer van der Walls interactions. The intralayer and interlayer C-$2p_z$ orbital hybridizations, respectively, create the gapless Dirac-cone structure [32] and the weak, but significant overlaps of valence and conduction bands [33]. The latter are responsible for the asymmetric energy spectra of valence holes and conduction electrons about the Fermi level. The low-lying $\pi$ electronic states are initiated from the K and H valleys [the inset of Fig. 2]. Furthermore, their state energies are, respectively, lower and higher than the Fermi energy in term of a weak energy dispersion, i.e., there exist the 3D free carriers of the hole and electron pockets. Very interesting, the other essential properties are easily further modulated by the external factors, such as, the intercalation-/de-intercalation- [34], temperature-, pressure-, and magnetic-field-enriched phenomena [35,36,37]. In addition, the systematic investigations on bulk graphite systems could be found review articles and books [38] . Most important, the well-behaved $\pi$ bands could be clearly identified the energy spectra along KMΓ and HLA with the whole widths more than 7 eV. The wide $\pi$-band widths are attributed to a close cooperation of the inralayer and interlayer carbon-$2p_z$ orbital hybridizations. On the other side, the $\sigma$ orbitals of [$2p_x$, $2p_y$,2s] could built two degenerate bands and one band below at the Γ point below -3 and -10 eVs, respectively. These features are mainly determined by the $\sigma$-electronic hopping integrals and ionization energies [39].

$AlCl_4^-$ -ion and $AlCl_4$ -molecule intercalations, which are, respectively, shown in Figs. 2(c)-(f) and Figs. 2(g)-(j), are able to create the diverse energy spectra and wave functions. The dramatic changes cover the variation of high-symmetry points, the creation of a lot energy subbands, the greatly enhanced asymmetry of occupied and unoccupied spectra about the Fermi level, the obvious reduce or enhancement of band overlaps [the diversified free carrier densities], the almost isotropic/highly anisotropic features near/away from $E_F=0$, the various energy dispersions with the different critical points, the frequently crossing & anti-crossing behaviors, the non-well-behaved $\pi$-$\sigma$-band widths, the carbon-, aluminum- &chloride-dominances at the different energy ranges [blue circles, red triangles and green squares, respectively]. Moreover, the $\pi$ & $\sigma$ electronic states could be easily identified from the original valleys, but not their whole band widths. After the chemical modifications, the enlarged Moire superlattices possess many atoms/ions in primitive unit cells [Fig. 1], so that the hexagonal first Brillouin zone in Fig. 2(a) is diminished quickly, especially for the low concentration cases. This leads to many valence and conductions with the smaller wave-vector ranges [40]. Due to the zone-folding effects [41], the low-lying electronic states are initiated from the Γ and A valleys [the K and H ones] under the cases of 1:18, 1:24 & 1:54 [1:32]. As to the ionic chemical environments, a pair of anisotropic valence and conduction bands across the Fermi level, which appears in a pristine Bernal graphite [Fig. 2(b)], is changed into an isotropic Dirac-cone structure of monolayer graphene. Furthermore, the energy spectra are dispersionless along the ΓA or KH directions. This clearly illustrates the semiconducting behaviors with a zero-band gap and density of state at $E_F=0$, simultaneously indicating the very weak carbon-intercalant-ion orbital hybridizations under the saturated atomic configurations. These graphite intercalation compounds are expected to present the lower electrical conductivities after the chemical reactions [42], However, they become outstanding merits in aluminum-ion transports and cathode intercalations/de-intercalations [43], Very interesting, the $\pi$-electronic dominance in the energy range of $E^{c,v} \leq 1.0$ eV also comes to exist in $AlCl_4$-molecule graphite intercalations, The obvious red shifts of $E_F$'s is revealed in any chemical cases. Electrons are largely transferred from carbon atoms to molecules, where the latter possess the larger affinities [44]. The strong p-type doping effects should be attributed to the significant carbon-molecule orbital hybridizations. In short, three types of band structures, semimetal, semiconductor and metal, respectively, arise from the interlayer van der Waals, carbon-saturated-ion and carbon-molecule interactions. Whether the similar phenomena could be found in other graphite intercalation compounds deserves a closer VASP simulation

The atom dominance, which corresponds to the spatial distribution probability of each wave-vector state, is clearly revealed in the specific energy ranges. It is determined by the intrinsic orbital hybridizations of chemical bonds. Most of electronic states in the whole energy spectrum is dominated by carbon atoms for the various $AlCl_4^-$-ion and $AlCl_4$-molecule intercalations [blue open circles in Figs. 2(c)- (f) and Figs. 2(g)-(j) for the $\pi$, $\sigma$ and $\pi^*$ energy subbands]. This is attributed to the dominating $\pi$, $\sigma$, carbon-intercalant bondings, since their initial state energies remain there and the frequent anti-crossing behaviors come to exist [45]. Specifically, aluminum atoms make observable contributions near $E^v \sim$ -4 and -6 eVs [green squares], suggesting the linking roles through a large ion/molecule structure. As for chloride atoms, the unusual roles are revealed as the weakly dispersive valence bands, at least four ones, below the Fermi level more than 1 eV [red triangles]. Their main features, energy, degeneracy, spacing and group velocity of valence subband, are very sensitive to the change of intercalant configuration. The partially flat subbands, with the zero velocities [the localized behaviors] frequently appear under the large-ion intercalations [Figs. 2(c)- (f)]. However, they might exhibit the observable modifications under the molecular cases [Figs. 2(g)- (j)]. The larger carrier mobility clearly indicates the more extensive charge distributions. The above-mentioned characteristics might be closely related to all the active chemical bondings, being supported by the further discussions in charge density distributions [Fig. 3] and van Hove singularities [Fig. 4].

The theoretical predictions of occupied electronic states below the Fermi level could be examined by the angle resolved photoemission spectroscopy [ARPES; 46], as discussed in Chap. 3.3 in detail. In general, it is very difficult to measure the $k_z$ - dependent energy spectra because of the destruction of momentum conservation through the surface boundary How to utilize the most important band features along KH, ML and ΓA would become a critical technique in identifying energy dispersions. The high-resolution ARPES measurement have successfully conducted on the semi-metallic energy bands in Bernal graphite, but not those of the rhombohedral and simple hexagonal graphites [ABC- and AA-stacked ones; 47,48]. According to the calculated results, the second and third systems, respectively, the lowest and highest free carrier densities, mainly owing to the symmetry of stacking configuration [49]. As to the $AlCl_4^-$-ion/ $AlCl_4$-molecule intercalation of stage-3/stage-4 graphite intercalation compounds, the observed occupied energy spectra are expected to exhibit the greatly diversified phenomena in terms of stacking-symmetry dependences, red shifts of the Fermi level, band overlaps [free carrier densities], strong energy dispersions, high anisotropies, and the characterizations of $\pi$ - and $\sigma$ - electronic energy spectra. In addition to these features, the VASP simulations on the stage-1 systems is able to provide the very useful information about the chloride- and aluminum-related valence bands. The further experimental examinations are very helpful to thoroughly clarify the intercalation/de-intercalation effects on electronic energy spectra and wave functions, as well as the intrinsic quasiparticle properties of orbital hybridizations. [50]

*The active orbitals hybridizations*

The spatial charge distributions [$\rho(r)s$] and their variations [$\Delta\rho(r)s$] after the chemical intercalations, as clearly shown in Fig. 3, are capable of providing very useful chemical pictures in fully comprehending the critical chemical bonds with the active orbital hybridizations. The [x, y]-top, [x, z]-side and [y, z]-side views fully illustrate the rich and unique intrinsic interactions: the C-C bonds in a honeycomb lattice, the interlayer C-Cl bonds, and the intra-ion/intra-molecule Al-Cl & Cl-Cl bonds. The significant chemical bondings agree with density of states [the merged van Hove sigularities in Fig. 4] and band structures [atom dominances in Fig. 2]. The strong evidence are thoroughly identified from $\rho(r)s$ and $\Delta\rho(r)$ s. First, the prominent $\sigma$ binding, which survives in a pristine Bernal is revealed as a very strong covalent between two neighboring carbon atoms [a super-high charge density by the red color on the [x, y] plane in Fig. 3(a); 51]. Furthermore, the $\pi$ bonding is characterized by the wave-like charge distribution $\rho(r)s$ and their variations $\Delta\rho(r)s$ due to the parallel significant $2p_z$-orbital hybridizations, as examined from $\rho(r)$ on both [x, z] and [y, z] planes [Fig. 3(a)]. Its distribution along the z-direction is somewhat extended by the interlayer van der Waals interactions [52]. Secondly, the significant carbon-intercalant couplings, corresponding to the large-ion cases [Figs. 3(b)-(e)], are directly reflected in the strongly anisotropic charge distributions $\rho(r)s$ in the 1st, 2nd and 3rd plots] and their variations $\Delta\rho(r)s$ near chloride atoms and the drastic changes between them, especially for the [x, z]- and [y, z]-plane projections. On the other side, the large-molecule intercalations, as indicated in Figs. 3(f)-(i), are able to greatly enhance charge density distributions in C-Cl bonds, since its atomic configuration belongs the non-closed-shell status. As for the Al-Cl and Cl-Cl bonds, they present the prominent bondings through the obvious distorted charge densities near chloride and aluminum atoms. In addition, the calculated results are very difficult to examine the existence of AL-C and Al-Al bonds. Furthermore, they cannot provide the enough information in examining the effects of the inter-ion/inter-molecule intercations. The similar analyses could be generalized to other multi-component graphite intercalation compounds, e.g., the active chemical bonds in $H_2SO_4$-[53], $HNO_3$-[54], and $FeCl_3$-[55]related systems.

According to the well definition of density of states, D(E) is expressed as the integration of the inverses of group velocities on the constant-energy configuration. For example, the 3D/2D/1D D's are greatly enriched by the various first derivatives of the gradient operations on electronic energy spectra and the specific-E integrations on the closed shells/circles/but the two discrete wave-vector points. A vanishing group velocity comes to exist while it corresponds to a critical point in the energy-wave-vector space. The singular integration function will lead to a special structure, namely, a van Hove singularity. The main features of singular structures, their forms, intensities, energies and numbers are very sensitive to the characteristics of distinct critical points and dimensionalities [56]. In general, the former are classified into the extreme, saddle and partially flat points, being clearly illustrated by the linear, parabolic, almost dispersionless and sombrero-shape energy dispersions of few-layer graphene systems [57]. When the orbital- & orbital-decomposed density of states are done for any condensed-matter systems, the various singular structures, with the prominent intensities, are available in determining the active orbital hybridizations of different chemical bonds [58]. This is based on their great enhancements through the emerged van Hove singularities [59]. As for the large-intercalant graphite intercalation compounds, there are a plenty of atom- & orbital-projected components. The very complicated results need to be analyzed in detail. [60].

For each intercalation case of $AlCl_4^-/AlCl_4$, there are one atom- and three orbital-decomposed density of states, being rather sufficient in providing the useful information about the active multi-/single-orbital hybridizations of distinct chemical bonds [61]. Very apparently, Fig. 4 show the rich and unique van Hove singularities mainly due to [C, Cl, Al] atoms and their significant orbitals. The magnitude of D(E) at the Fermi level represents the characteristics of free carriers. Bernal graphite and ion intercalation systems have the low values at $E_F=0$, as well as band structures, respectively, suggesting the semi-metallic and semiconducting behaviors. However, each large-molecule case exhibits a finite value there. Most important, the difference between the Fermi level and the featured energy with the smallest density of states could be regarded its red shift [the details in Table 2]. Furthermore, this covered area just corresponds to the total free carriers per unit cell after the obvious charge transfers from carbon to chloride [Fig. 3]. The stronger affinity of the latter is responsible for the p-type doping effects [the free valence holes; 62]. This quantity is deduced to be proportional to the $AlCl_4$-molecule concentration. In addition, the Fermi-momentum states of electronic spectra [Figs. 2(g)-(j)] are not reliable in evaluating the transferred valence hole density in the presence of complicated zone-folding effects.

The significant chemical bonds and their active multi-/single-orbital hybridizations are further achieved from the delicate analyses, covering all the separated and merged van Hove singularities [Fig. 4]. The concise physical and chemical pictures are also supported by electronic energy spectra [Figs. 2] and spatial charge distributions [Fig. 3]. Both $AlCl_4^-$ and $AlCl_4$ graphite intercalation compounds possess the intralyer carbon-carbon, interlayer carbon-intercalant, and intra-/inter-intercalant interactions, respectively, leading to the C-C, C-Cl, and Al-Cl & Cl-Cl bonds. However, the observable evidence of merged van Hove singularities are absent for Al-C and Al-Al. The prominent chemical bondings are thoroughly illustrated as follows. Since $AlCl_4^-$ has a closed 0 shell atomic configuration, each graphitic sheet recovers to a pure honeycomb lattice. The $\pi$- and $\sigma$-electronic spectra are well separated from each other [all PDOS of C cases in Fig. 4]. Furthermore, the former [the pink curves] and the latter [the red, blue and black curves] are, respectively, characterized by the initial/prominent structures at ~ -2.0 eV and ~ -3.12 / -6,15 eV. In addition, two strong peaks mainly arise from the saddle points of valence $\pi$ and $\sigma$ bands. Apparently, $\pi$ and $sp^2$ bondings are orthogonal to each other and thus survive in C-C bonds. The interlayer C-Cl bonds are revealed as the multi-orbital hybridizations of $[2p_x, 2p_y, 2p_z]-[3p_x, 3p_y, 3p_z]$ through the emerged structures within -6.1 eV≤E≤-1.8eV. As for Al-Cl/Cl-Cl bonds, the obvious four-orbital hybridizations of $[3s, 3p_x, 3p_y, 3p_z]-[3s, 3p_x, 3p_y, 3p_z]/[3s, 3p_x, 3p_y, 3p_z]-[2s, 2p_x, 2p_y, 2p_z]$ are clarified from the van Hove singularities at -4.0eV and -5.9eV. Such unusual results are due to the fact 3s &$[3p_x,3p_y,3p_z]$-decomposed, respectively, appear the same and different energies for Al and Cl. The similar features of density of states could be found in the molecular intercalation cases, while the red-shift phenomena are created by the very strong p-type doping effects. That is, the important C-Cl bonds can enrich the valence van Hove singularities near the Fermi level. This is consistent with more charge variations between honeycomb lattice and intercalant layer. In addition, it is almost mission impossible to investigate the inter-ion and inter-molecule interactions from the van Hove singularities. [63]

**Conclusions**

In summary, Bernal graphite, stage-1 $AlCl_4^-$-ion and $AlCl_4$-molecule graphite intercalation compounds exhibit the diverse quasiparticle behaviors under the distinct orbital hybridizations of intralayer and interlayer chemical bonds. The effects, which are due to the significant van der Waals interactions, the rich intercalations of closed-shell ions, and the strong charge transfers, are responsible for the rich and unique properties. They are directly reflected in the distinct crystal symmetries [the periodical configurations perpendicular/on the (x, y) plane], the largely enhanced Moire superlattices, the existence of many valence & conduction energy subbands [the zone-folding effect], the variations about the initial high-symmetry valleys, the largely enhanced asymmetry of hole and electron spectra about the Fermi level, the vanishing band overlaps/the p-type doping effect [the zero/finite free carrier densities after ion/molecule interactions], the almost isotropic/highly anisotropic features close/away from the Fermi level, the various energy dependences at the distinct critical points, the frequent band crossings & mixings [the complicated orbital hybridizations], the non-well-defined $\pi$- & $\sigma$- electronic band widths, the carbon-, aluminum- & chloride-determined energy spectra at the distinct energy ranges. The active multi-/single-orbital hybridizations of C-C/C-Cl/Al-Cl/Cl-Cl chemical bonds are identified to be [2s, $2p_x$, $2p_y$]-[2s, $2p_x$, $2p_y$] & $2p_z$-$2p_z$/[2s, $2p_x$, $2p_y$, $2p_z$]-[3s, $3p_x$, $3p_y$, $3p_z$]/[3s, $3p_x$, $3p_y$, $3p_z$]-[3s, $3p_x$, $3p_y$, $3p_z$]/[3s, $3p_x$, $3p_y$, $3p_z$]-[3s, $3p_x$, $3p_y$, $3p_z$]. However, the observable evidence is very difficult to examine for C-Al and Al-Al bonds. The up-to-date experiments only verify the observation of stage-3 and stage-4 cases [64][65], in which the molecular dynamics could examine the existence of stage-1 and stage-2 configurations. Concerning the complicated intercalation and de-intercalation processes, their critical roles on the large-ion transports within the high-performance batteries require the systematic investigations, [66] especially for the development of theoretical frameworks.

**Acknowledgements**

This work is supported by Taiwan Ministry of Science and Technology under grant number MOST 108-2112-M-006-016-MY3 and MOST 109-2124-M-006-001.

**References**

[1] Sutter P, Sadowski JT and Sutter EA 2010 Chemistry under Cover : Tuning Metal-Graphene Interaction by Reactive Intercalation. *Journal of the American Chemical Society* .132, 8175-8179.

[2] Dale A, C Brownson and Craig E Banks 2010 Graphene electrochemistry: an overview of potential applications. *Analyst* 135, 2768-2778.

[3] CY Lin, JY Wu, YJ Ou, YH Chiu and MF Lin 2015 Magneto-electronic properties of multilayer graphenes. *Physical Chemistry Chemical Physics*, 17(39), 26008.

[4] RozpLlocha F, Patyk J and Stankowski J 2007 Graphenes Bonding Forces in Graphite. *Institute of Physics, Nicolaus Copernicus UniversityGrudzi* 5/7, 87-100.

[5] Brownson DAC and Banks CE 2010 Graphene electrochemistry: an overview of potential applications . *Analyst* 135, 2768-2778.

[6] Hu M, Dong X, Wu YJ, Liu LY and Zhao ZS 2018 Low-energy 3D sp(2) carbons with versatileproperties beyond graphite and graphene. *Dalton Transactions 47, 6233-6239* .

[7] Konschuh S, Gmitra M and Fabian J 2010 Tight-binding theory of the spin-orbit coupling in graphene. *Physical Review B* 82, 245412.

[8] Huang JR, Lin JY, Chen BH and Tsai MH 2008 Structural and electronic properties of few-layer graphenes from first-principles. *Physica Status Solid B-Basic Solid State Physics* 245, 136-141.

[9] de Andres PL, Ramirez R and Verges JA 2008 Strong covalent bonding between two graphene layers. *Physical Review B* 77, 045403.

[10] Abergel DSL, Apalkov V, Berashevich J, Ziegler K and Chakraborty T 2010 Properties of graphene: a theoretical perspective. *Advances In Physics* 59, 261-482.

[11] Lee SH, Chiu CW and Lin MF 2010 Deformation effects on electronic structures of bilayer graphenes.*Physica E-low-Dimensional Systems &Nanost- Ructures 42,* 732-735.

[12] Briggs N, Gebeyehu ZM, Vera A, Zhao T, Wang K, Duran AD, Bersch B, Bowen T, Knappenberger KL and Robinson JA.2019 Epitaxial graphene/silicon carbide intercalation: a minireview on graphene modulation and unique 2D materials. *Nanoscale* 11, 15440-15447.

[13] Silva CC, Cai JQ, Jolie W, Dombrowski D, zum Hagen FHF, Martinez-Galera AJ, Schlueter C, Lee TL and Busse C 2019 Lifting Epitaxial Graphene by Intercalation of Alkali Metals.*Journal of Physical Chemistry C 123*, 3712-13719.

[14] Yoo H, Engelke R, Carr S, Fang SA, Zhang K, Cazeaux P, Sung SH, Hoyden R, Tsen AW and Taniguchi T 2019 Atomic and electronic reconstruction at the van der Waals interface in twisted bilayer graphene. *Nature Materlals 18,* 448.

[15] Kaneko T and Saito R 2017 First-principles study on interlayer state in alkali and alkaline earth metal atoms intercalated bilayer graphene. *Surface Science 665 1-9.*

[16] Silva CC, Cai JQ, Jolie W, Dombrowski D, zum Hagen FHF, Martinez-Galera, AJ, Schlueter C, Lee Tl and Busse C 2019 Lifting Epitaxial Graphene by Intercalation of Alkali Metals. *Journal Of Physical Chemistry C 123,* 13712-13719.

[17] J. H. Ho, C. L. Lu, C. C. Hwang, C. P. Chang, and M. F. Lin 2006 Coulomb excitations in AA- and AB-stacked bilayer graphites. *Physical Review B*, 74(8), 085406.

[18] Chun-Jern Pana, Chunze Yuana, Guanzhou Zhua, Qian Zhangc, Chen-Jui Huangb, Meng-Chang, LindMichaelAngella, Bing-JoeHwangb, Payam Kaghazchic and Hongjie Daia 2018 An operando X-ray diffraction study of chloroaluminate anion-graphite intercalation in aluminum batteries. *Pnas* 115, 22.

[19] Preeti Bhauriyal, Arup Mahata and Biswarup Pathak 2017 The staging mechanism of AlCl4 inter & calation in a graphite electrode for an aluminium-ion battery. *Phys.Chem.Chem.Phys* 19, 7980.

[20] MS Wua, B Xua, LQ Chenb and CY Ouyanga. 2016 Geometry and fast diffusion of AlCl4 cluster intercalated in graphite. *Electrochimica Acta* 195, 158–165.

[21] Do.TN, Chang.CP, Shih.PH, Wu.JY and Lin.MF 2017 Stacking enrichedmagneto transport properties of few-layer graphenes. *Physical Chemistry Chemical Physics*, 19, 29525-29533.

[22] Leon A and Pacheco M 2011 Electronic and dynamics properties of a molecular wire of graphane nanoclusters. *Physics Letters A* 375, 4190-4197.

[23] Mounet N and Marzari N 2005 First-principles determination of the structural, vibrational and thermodynamic properties of diamond, graphite, and derivatives. *Physical Review B* 71, 205214.

[24] Yurui Gao, Chongqin Zhu, Zheng Zheng Chen and Gang Lu 2017 Understanding Ultrafast Rechargeable Aluminum-Ion Battery from First-Principles. *J. Phys. Chem. C* 121 7131–7138.

[25] M. F. Lin, C. S. Huang, and D. S. Chuu 1997 Plasmons in graphite and stage-1 graphite intercalation compounds. *Phys. Rev. B* 55, 13961.

[26] Alder BJ and Wainwright TE 1959 Studies In Molecular Dynamics .1. General Method . *Ournal Of Chemical Phtsics* 31**,** 459-466.

[27] Di-Yan Wang, Shao-Ku Huang, Hsiang-Ju Liao, Yu-Mei Chen, ShengWen Wang, Yu-Ting Kao, Ji-Yao An, Yi-Cheng Lee, Cheng-Hao Chuang, Yu-Cheng Huang, Ying-Rui Lu, Hong-Ji Lin, Hung-Lung Chou, Chun-Wei Chen, Ying-Huang Lai and Chung-LiDong 2019 Insights into dynamic molecular intercalation mechanism for Al-C battery by operando synchrotron X-ray techniques.*Carbon* 146 **,** 528-534.

[28] C. Y. Lin, R. B. Chen, Y. H. Ho and M. F. Lin.2018 Electronic and optical properties of graphite-related systems. *CRC Press, Boca Raton, Florida.*

[29] Malko D, Neiss C, Vines F and Gorling A 2012 Competition for Graphene: Graphynes with

Direction-Dependent Dirac Cones. *Physical Revier Letters* 108, 086804.
[30] Castro Neto AH, Guinea F, Peres NMR, Novoselov KS and Geim AK 2009 The electronic properties of graphene. *Reviews Of Modern Physica* 81,109-162.
[31] Abergel D, Apalkov V, Berashevich J, Ziegler K andChakraborty T 2010 Properties of graphene: a theoretical perspective. *Advances In Physics* 59,261-482.
[32] Wu CJ and DasSarma S 2008 px,y-orbital counterpart of graphene: Cold atoms in the honeycomb optical lattice . *Physical Review B* .77,23.
[33] Sprinkle M, Siegel D, Hu Y, Hicks J, Tejeda A, Taleb-Ibrahimi, Le Fevre, P, Bertran F, Vizzini S and Enriquez H 2009 First Direct Observation of a Nearly Ideal Graphene Band Structure. *Physical Review Letters* .103,226803.
[34] Dresselhaus Ms and Dresselhaus G 2002 Intercalation compounds of graphite. *Advances In Physica* 51,1-186.
[35] C. Y., J. Y. Wu, Y. H. Chiu, and M. F. Lin 2014 Stacking-dependent magneto-electronic properties in multilayer graphenes. *Physical Review B* 90(20),205434.
[36] J. Y. Wu, S. C. Chen, Oleksiy Roslyak, Godfrey Gumbs, and M. F. Lin.2011 Plasma Excitations in Graphene: Their Spectral Intensity and Temperature Dependence in Magnetic Field. *ACS NANO* 5(2), 1026.
[37] Goerbig MO 2011 Electronic properties of graphene in a strong magnetic field. *Reviews Of Modern Physics* 83,1193-1243.
[38] Shih-Yang Lin, Shen-Lin Chang, Feng-Lin Shyu, Jian-Ming Lu, and Ming-Fa Lin 2015 Feature-rich electronic properties in graphene ripples. *Carbon* 86,207-216.
[39] Shallcross S, Sharma S, Kandelaki E and Pankratov OA 2010 Electronic structure of turbostratic graphene. *Physical Review B* 81,165105.
[40] Reich S, Maultzsch J, Thomsen C and Ordejon P 2002 Tight-binding description of graphene. *Physical Review B* 66,035412.
[41] Lin Y, Chen G, Sadowski JT, Li YZ, Tenney SA, Dadap JI, Hybertsen MS and Osgood RM 2019 Observation of intercalation-driven zone folding in quasi-free-standing graphene energy bands *Physical Review B* 99,035428.
[42] Loh KP, Bao QL, Ang PK and Yang JX 2010 The chemistry of graphene. *Journal Of Materlals Chemistry* 20,2277-2289.
[43] Jung SC, Kang YJ, Yoo DJ, Choi Jw and Han YK 2016 Flexible Few-Layered Graphene for the Ultrafast Rechargeable Aluminum-Ion Battery. *Journal Of Physical Chemistry C* 120,13384-13389.
[44] Qianpeng Wang, Daye Zheng, Lixin He and Xinguo Ren 2019 Cooperative Effect in a Graphite Intercalation Compound: Enhanced Mobility of AlCl4 in the Graphite Cathode of Aluminum-Ion Batteries. *Physical Review Applied* 12 , 044060.
[45] T Kiss1, Shimojima1 K, Ishizaka1 A, Chainani T, Togashi T, Kanai1 X, Y Wang C, T Chen, S Watanabe and S Shin 2008 A versatile system for ultrahigh resolution, low temperature, and polarization dependent Laser-angle-resolved photoemission spectroscopy. *Review of Scientific Instruments* 79 ,023106.
[46] C. Y. Lin, C. H. Yang, C. W. Chiu, H. C. Chung, S. Y. Lin, and M. F. Lin. 2021 Many-particle interactions in carbon nanotubes. *IOP Concise Physics, San Raefel, CA, USA: Morgan & Claypool Publishers, in print.*
[47] Hwang EH and Das Sarma S 2008 Quasiparticle spectral function in doped graphene: Electron-electron interaction effects in Arpes *Physical Review B* 77,081412.
[48] C. Y. Lin, M. H. Lee, and M. F. Lin. 2018 Coulomb excitations in trilayer ABC-stacked graphene. *Physical Review B Rapid communication*, 98(4),041408.
[49] Tasaki K 2014 Density Functional Theory Study on Structural and Energetic Characteristics of Graphite Intercalation Compounds. *Journal Of Physical Chemistry C* 118,1443-1450.
[50] Li. WB, Lin SY, Tran NTT, Lin MF, and Lin KI. 2020 Essential geometric and electronic properties in stage-n graphite alkali-metal-intercalation compounds. *RSC Advances*,10,23573-23581.
[51] Geim AK and Mac Donald AH 2007 Graphene: Exploring carbon flatland. *Physics Today* 60,35-41.
[52] Duong DLF, Yun Sj and Lee YH. 2017 van der Waals Layered Materials: Opportunities and Challenges. *Acs Nano* 11,11803-11830.
[53] Hiroshi Shioyama, and Rokuro Fujii 1987 Electrochemical reactions of stage 1 sulfuric acid—Graphite intercalation compound. *Carbon* 25,771-774.
[54] NE Sorokina, NV Maksimova, AV Nikitin, ON Shornikova , and VV Avdeev 2001 Synthesis of Intercalation Compounds in the Graphite–$HNO_3$–$H_3PO_4$ System. *Inorganic Materials*,37, 584–590.
[55] Da Zhan, Li Sun, Zhen Hua Ni, Lei Liu, Xiao Feng Fan, Ying ying Wang, Ting Yu, Yeng Ming Lam, Wei Huang, Ze Xiang Shen 2010 $FeCl_3$-Based Few-Layer Graphene Intercalation Compounds: Single Linear Dispersion Electronic Band Structure and Strong Charge Transfer Doping. *Advanced Functionnal Materials.*
[56] Ro senzweig P, Karakachian H, Marchenko D, Kuster K, and Starke U. 2020 Overdoping Graphene Beyond the van Hove Singularity. *Physical Review Letters*,125 ,176403.
[57] Bostwick A, Ohta T, Seyller T, Horn K and Rotenberg E 2007 Quasiparticle dynamics in graphene. *Nature Physics* 3,36-40.
[58] Dresselhaus MS, and Dresselhaus G 2002 Intercalation compounds of graphite. *Advances in Physics* 51,1-186.
[59] Ro senzweig P, Karakachian H, Marchenko D, Kuster K and Starke U 2020 Overdoping Graphene Beyond the van Hove Singularity. *Physical Review Letters* 125,176403.
[60] Dresselhaus MS and Dresselhaus G 2002 Intercalation compounds of graphite. *Advances In Physics* 51,1-186.
[61] Tasaki k 2014 Density Functional Theory Study on

Structural and Energetic Characteristics of Graphite Intercalation Compounds *Journal Of Physical Chemistry C* 118,1443-1450.

[62] Liu HT,Liu YQ and Zhu DB 2011 Chemical doping of graphene . *Journal Of Materlals Chemistry* 21,3335-3345.

[63] Li GH,Luican A ,dos Santos JMBL ,Castro Neto, Reina, A ,Kong J and Andrei EY 2010 Observation of Van Hove singularities in twisted graphene layers. *Nature Physics* 6,109-113.

[64] Wang DY , Huang SK , Liao HJ , Chen YM ,Wang SW ,Kao YT ,An JY ,Lee YC Chuang CH and Huang YC 2019 Insights into dynamic molecular intercalation mechanism for Al-C battery by operando synchrotron X-ray techniques *Carbon* 146,528-534.

[65] Childress AS,Parajuli P , Zhu JY , Podila R and Rao AM 2017 A Raman spectroscopic study of graphene cathodes in high-performance aluminum ion batteries. *Biotechnology Advances* 29,189-198.

[66] Childress AS, Parajuli P, Zhu JY, Podila R, and Rao AM 2017 A Raman spectroscopic study of graphene cathodes in high-performance aluminum ion batteries. *Biotechnology Advances*, 29 ,189-198.

| | Concentration (AlCl4 / C) | Layer-Distance (Å) | C-C Bond-length (Å) | Al-Cl Bond-length (Å) | Al-Cl Bond-angel (°) |
|---|---|---|---|---|---|
| primitive | | 3.35 | 1.42 | 2.159 | 109.0 |
| 1∶54 | 1.85% | 10.65 | 1.424 | 2.165 | 108.01 |
| 1∶32 | 3.12% | 10.81 | 1.426 | 2.163 | 107.94 |
| 1∶24 | 4.16% | 11.29 | 1.427 | 2.162 | 107.46 |
| 1∶18 | 5.55% | 11.30 | 1.428 | 2.160 | 107.19 |

Table 1: The optimal geometric structures of $AlCl_4^-$-graphite intercalation compounds with the various concentrations: (a) 1:18, (b) 1:24, (c) 1:32 and (d) 1: 54 for the ration of $AlCl_4^-$ and C.

| | Concentration (AlCl4 / C) | Layer Distance (Å) | C-C Bond-length (Å) | Al-Cl Bond-length (Å) | Al-Cl Bond-angel (°) | Blue shifts (eV) |
|---|---|---|---|---|---|---|
| primitive | | 3.35 | 1.42 | 2.159 | 109.27 | |
| 1∶54 | 1.85% | 8.80 | 1.421 | 2.165 | 113.03 | 0.978 |
| 1∶32 | 3.12% | 8.81 | 1.422 | 2.163 | 112.00 | 0.911 |
| 1∶24 | 4.16% | 8.78 | 1.421 | 2.162 | 111.45 | 0.862 |
| 1∶18 | 5.55% | 8.77 | 1.422 | 2.160 | 111.7 | 0.753 |

Table 2: The similar results in Table 1, but illustrated for AlCl$_4$-graphite intercalation compounds. The blue shifts of the Fermi levels under the various concentrations are also shown for the charge transfer effects.

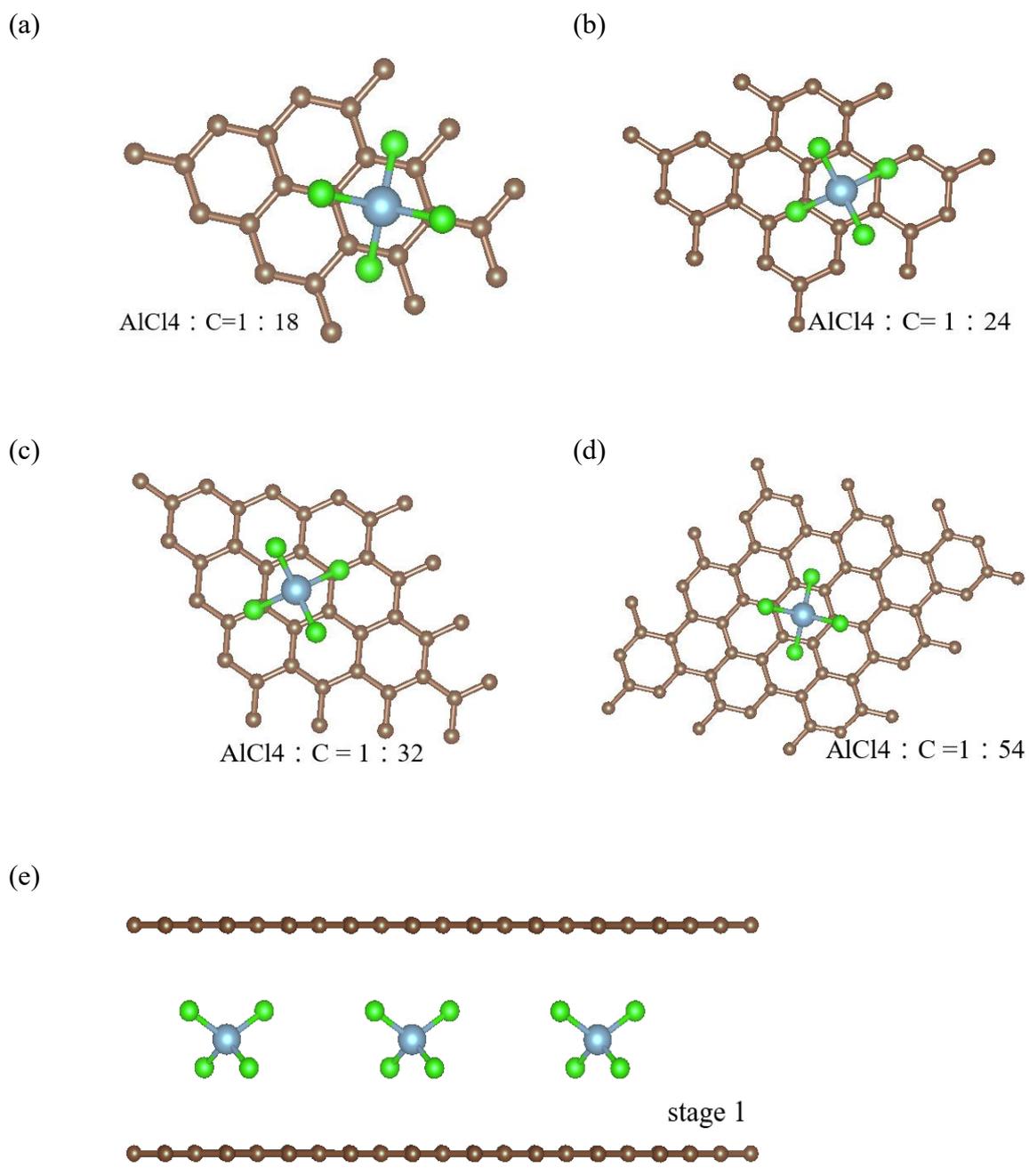

Figure 1: The optimal geometric structures of $AlCl_4^-/AlCl_4$ graphite intercalation compounds with the various concentrations: (a) 1:18, (b) 1:24, (c) 1:32 and (d) 1: 54 for the ration of $AlCl_4^-/AlCl_4$ and C, and (e) the side views are shown.

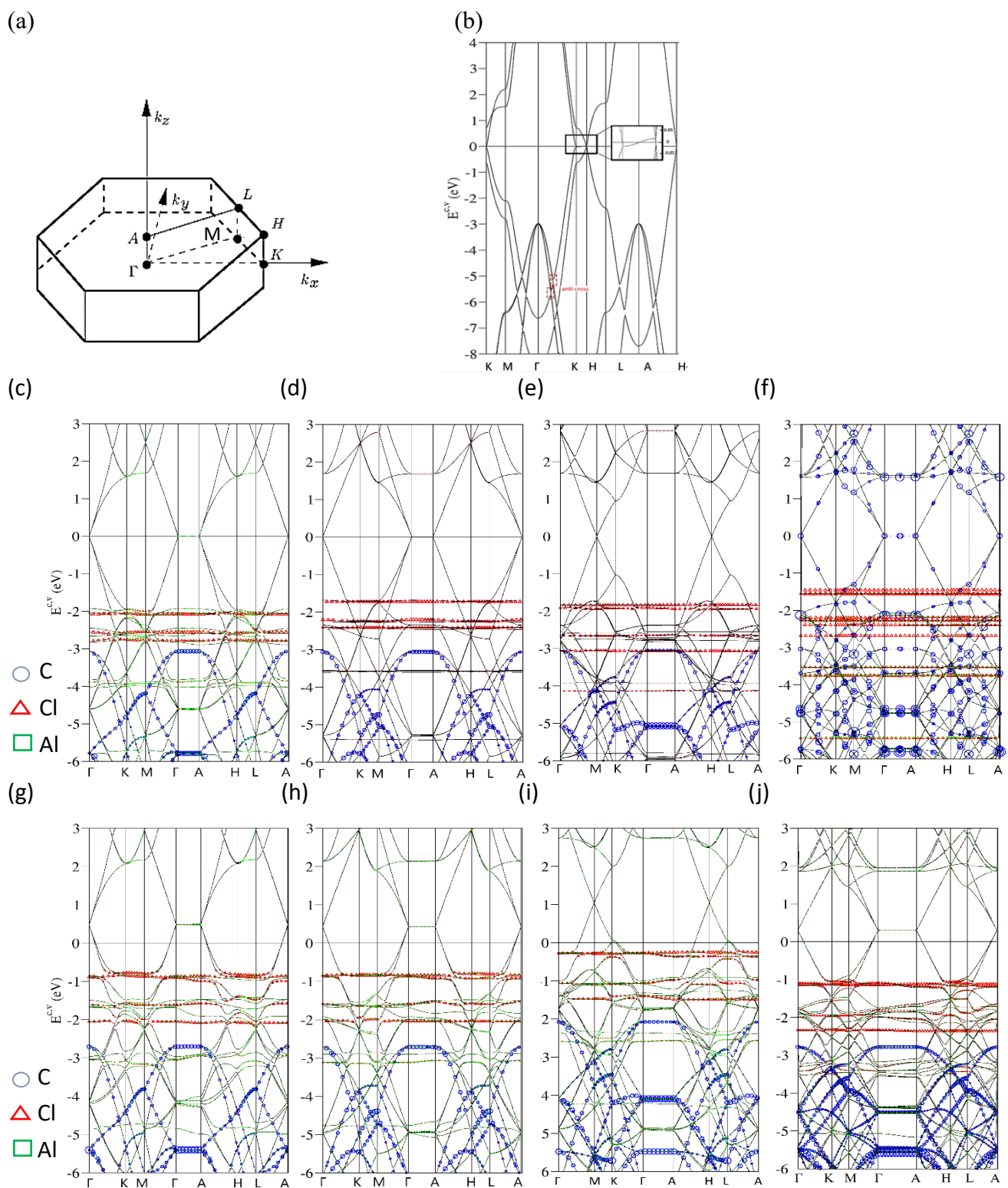

Figure 2: Band structures of $AlCl_4^-$/$AlCl_4$-related graphite intercalation compounds (a) along the high-symmetry points within the first Brillouin zone under the various cases: (b) a pristine system, (c)/(g) 1:18, (d)/(h) 1:24, (e)/(i) 1:32 and (f)/(j) 1: 54.

(a)

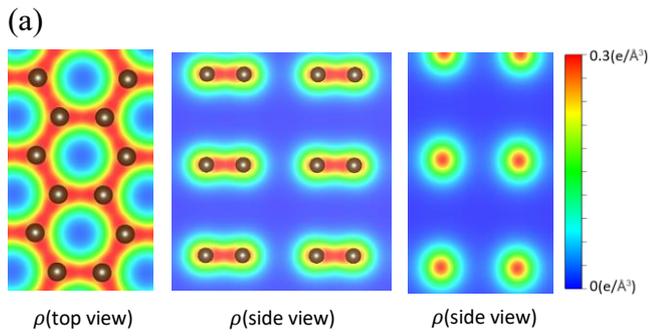

ρ(top view)　　ρ(side view)　　ρ(side view)

(b)

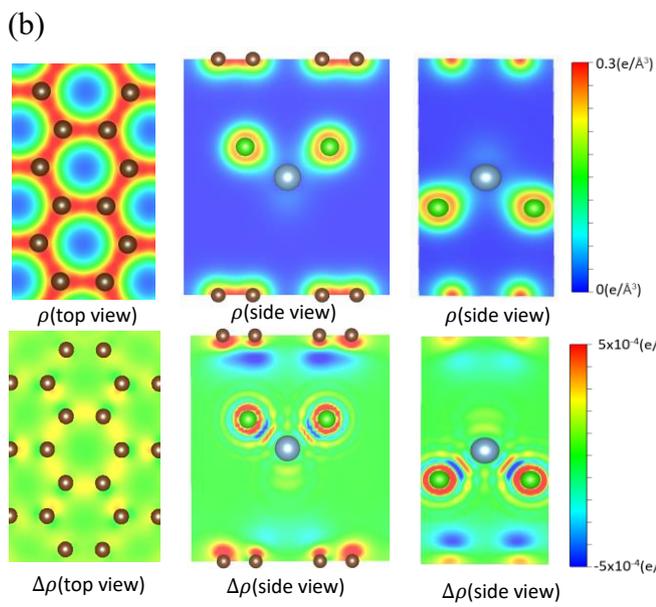

ρ(top view)　　ρ(side view)　　ρ(side view)

Δρ(top view)　　Δρ(side view)　　Δρ(side view)

(c)

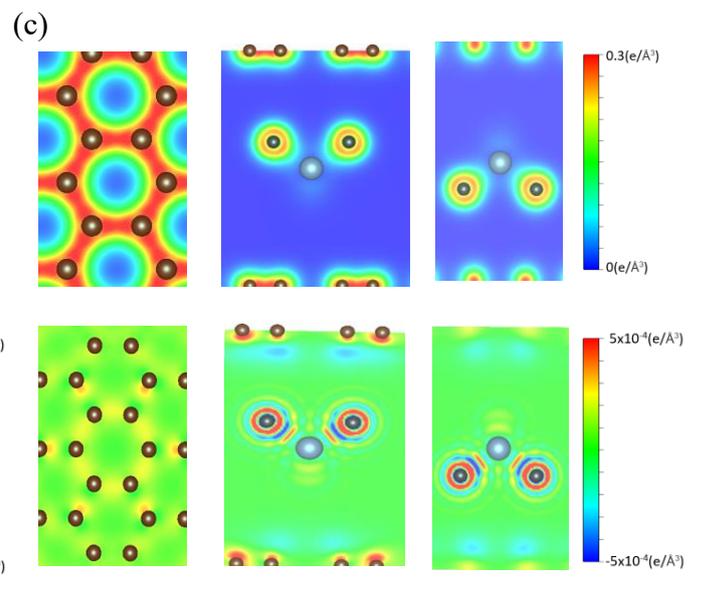

(d)

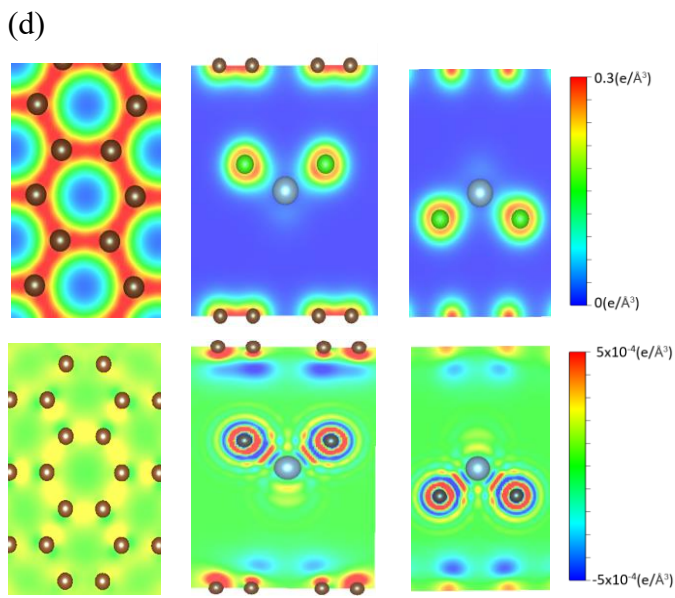

(e)

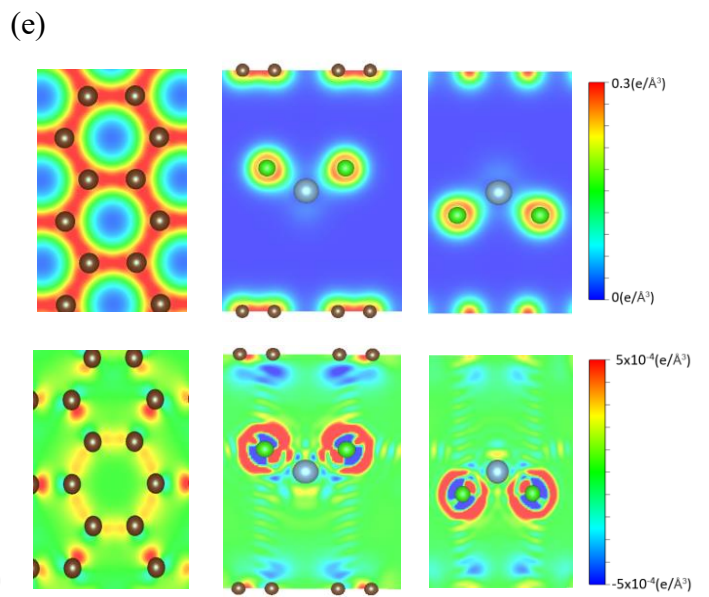

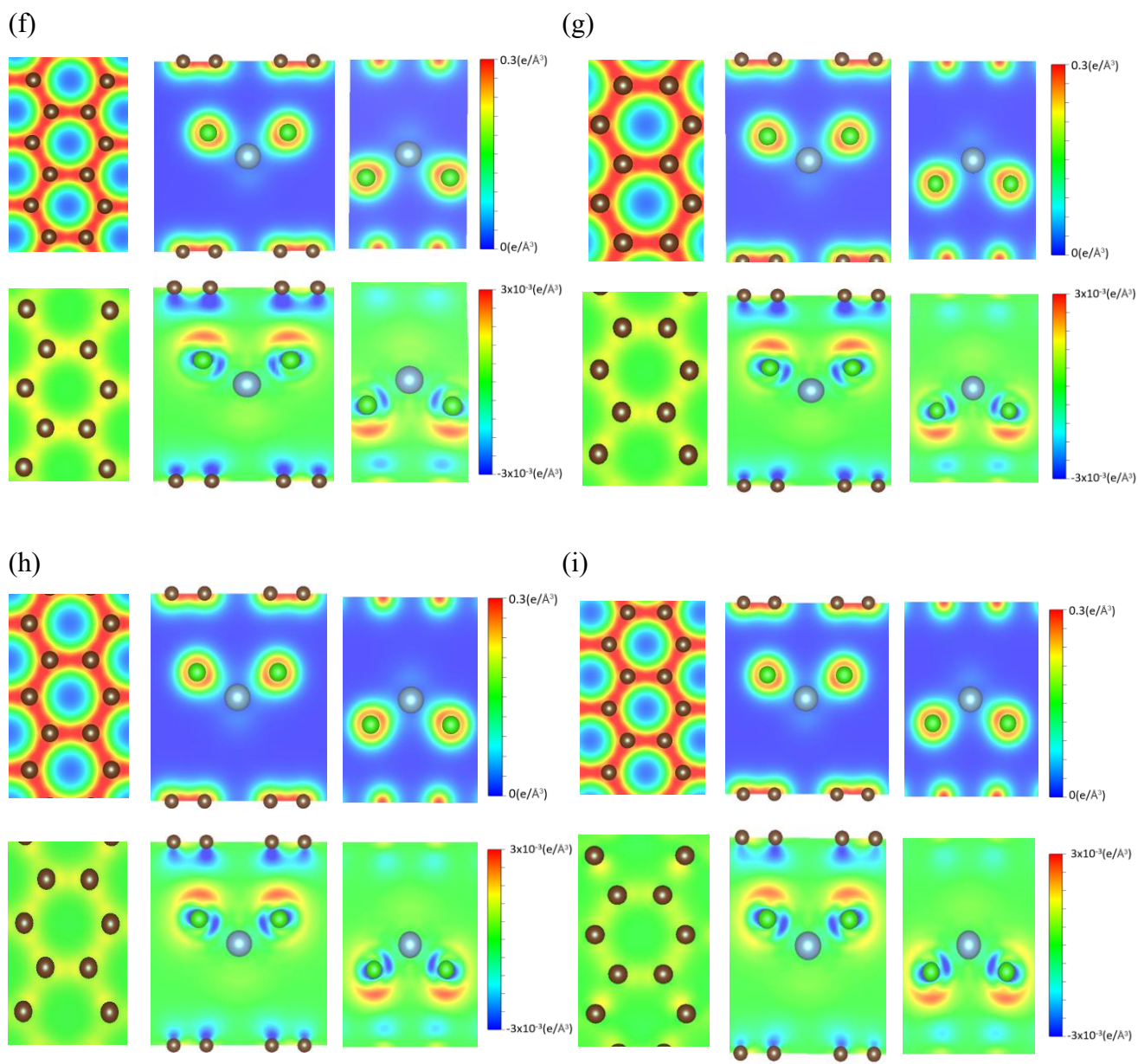

Figure 3: The spatial charge density distributions and their changes after $AlCl_4^-/AlCl_4$-ion intercalation into graphite under the distinct chemical cases: (a) a pristine system, (b)/(f) 1:18, (c)/(g) 1:24, (d)/(h) 1:32 and (e)/(i) 1: 54, with the top- and side views [the (x, y)-, (x, z)- and (y, z)-plane projections].

(a)

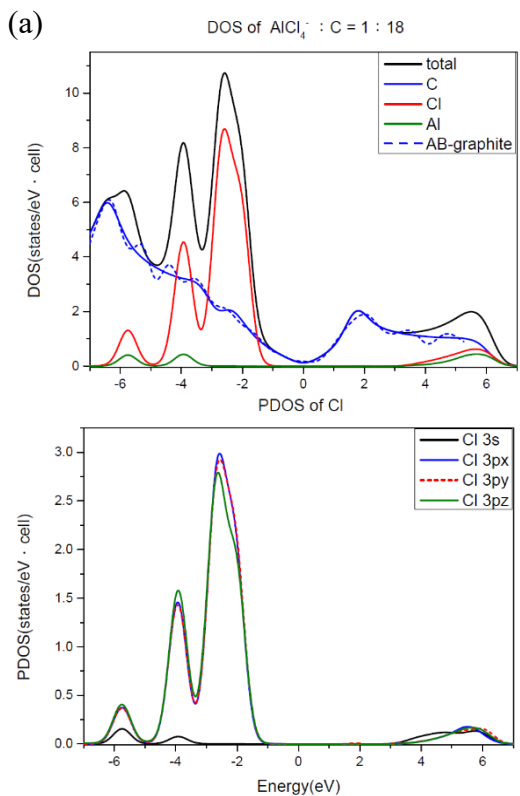
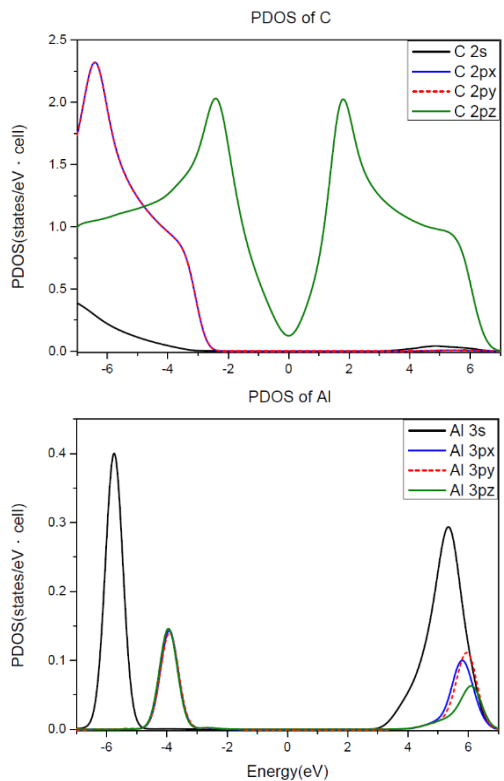

(b)

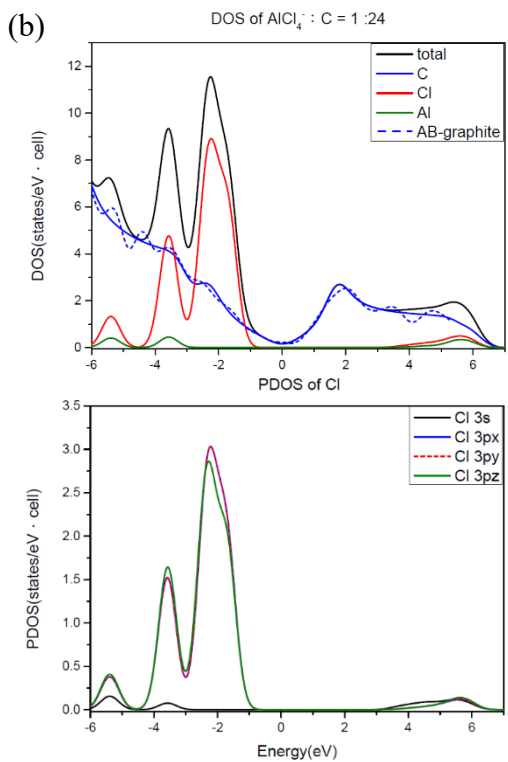
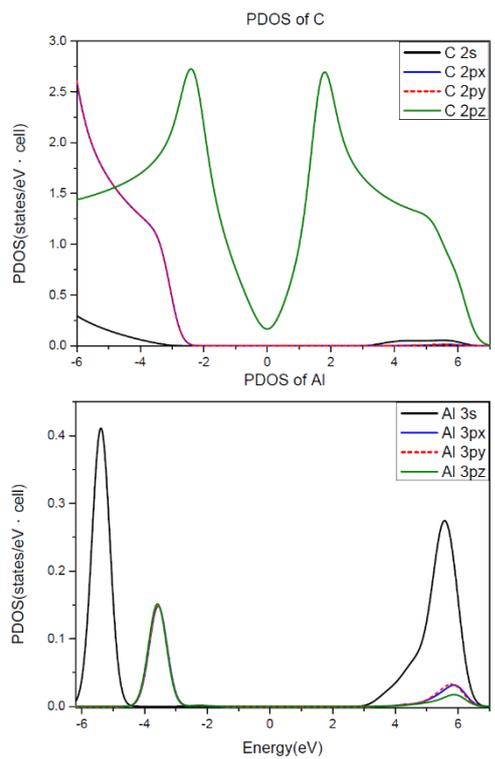

(c)

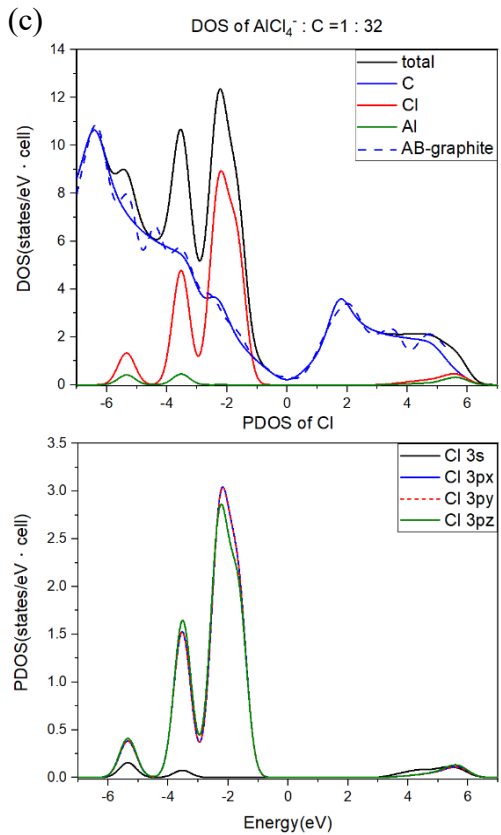
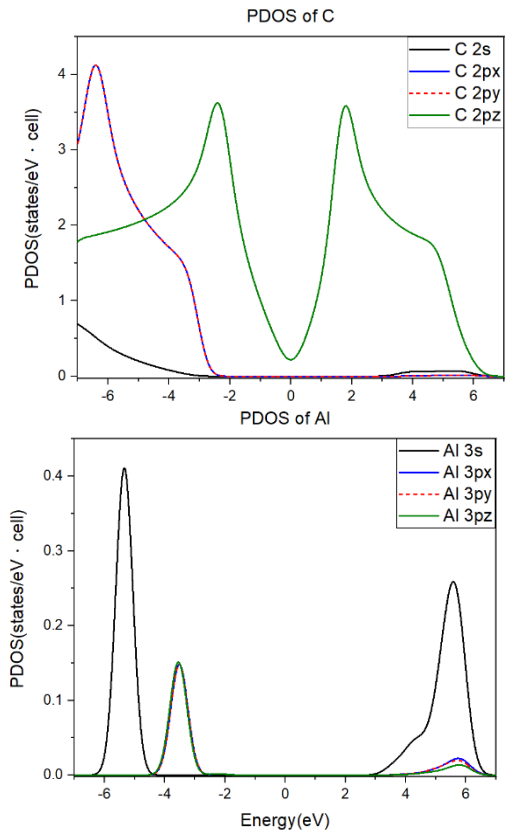

(d)

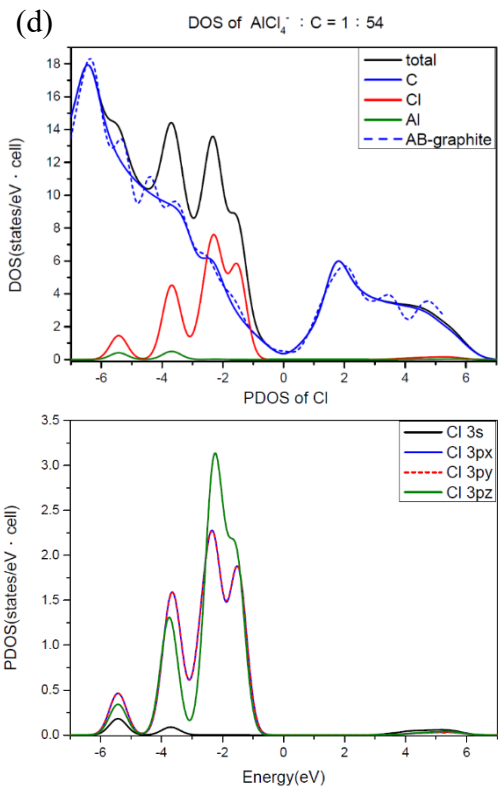
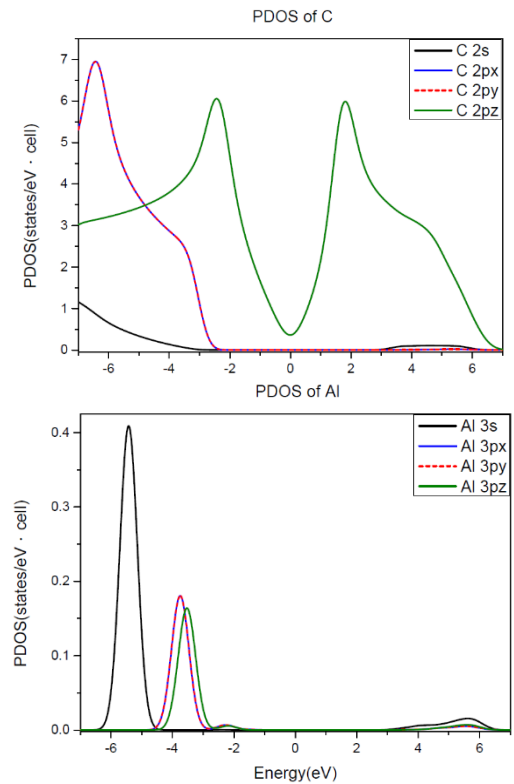

(e)

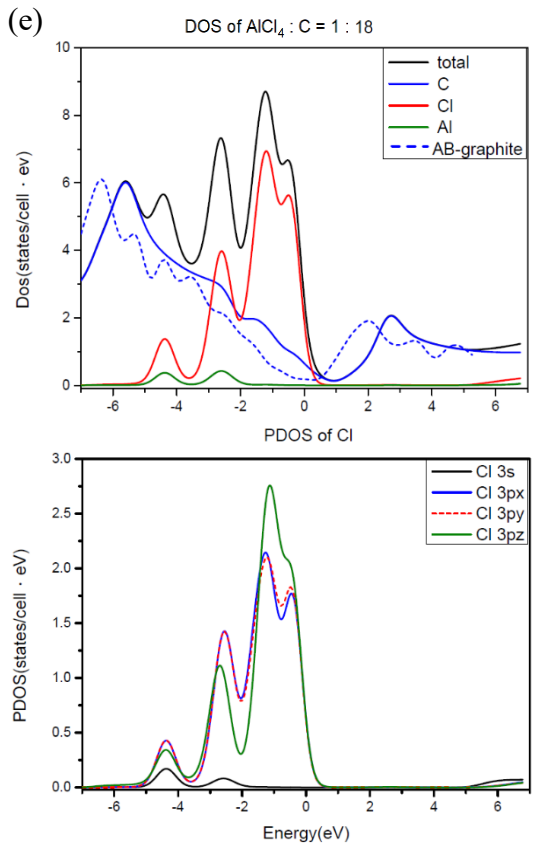
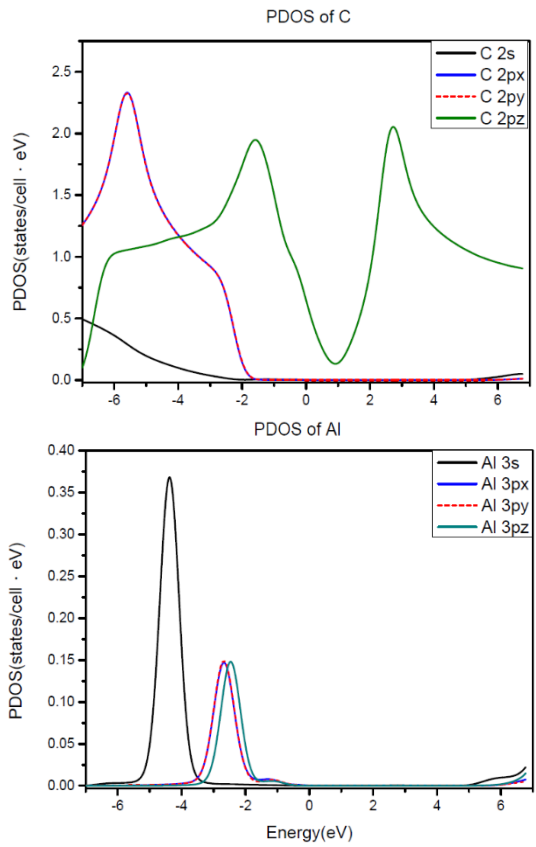

(f)

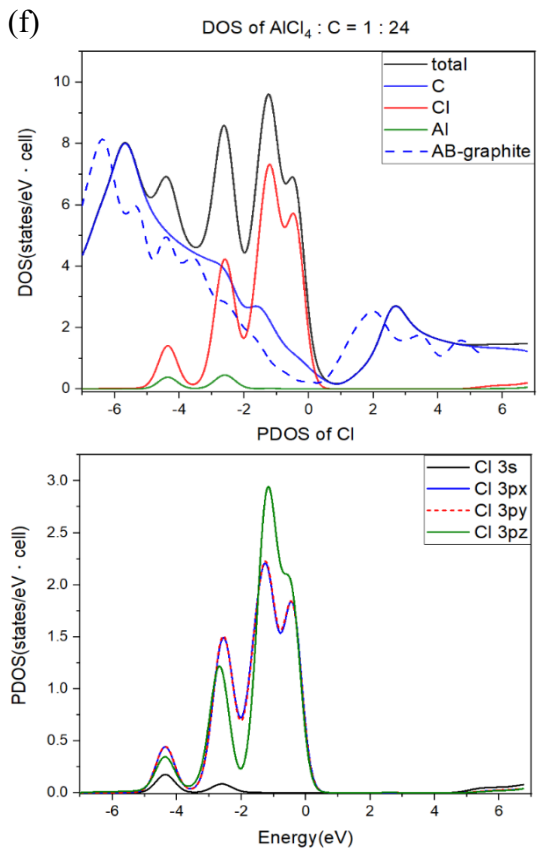
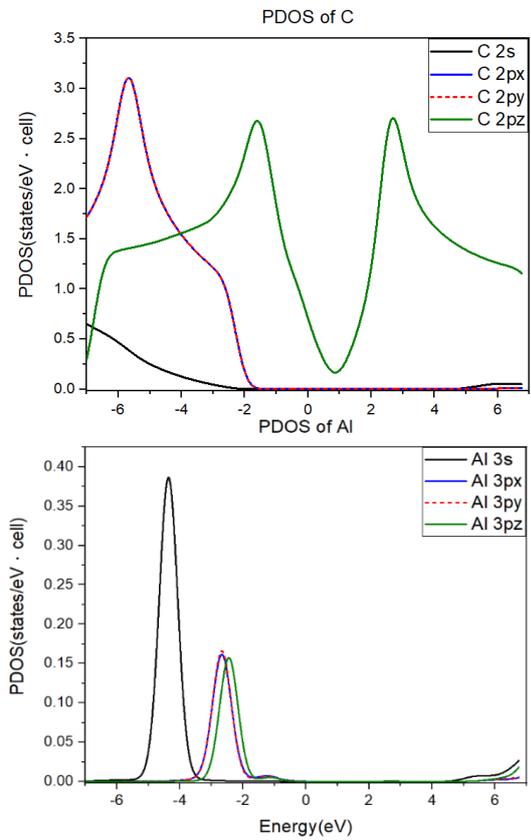

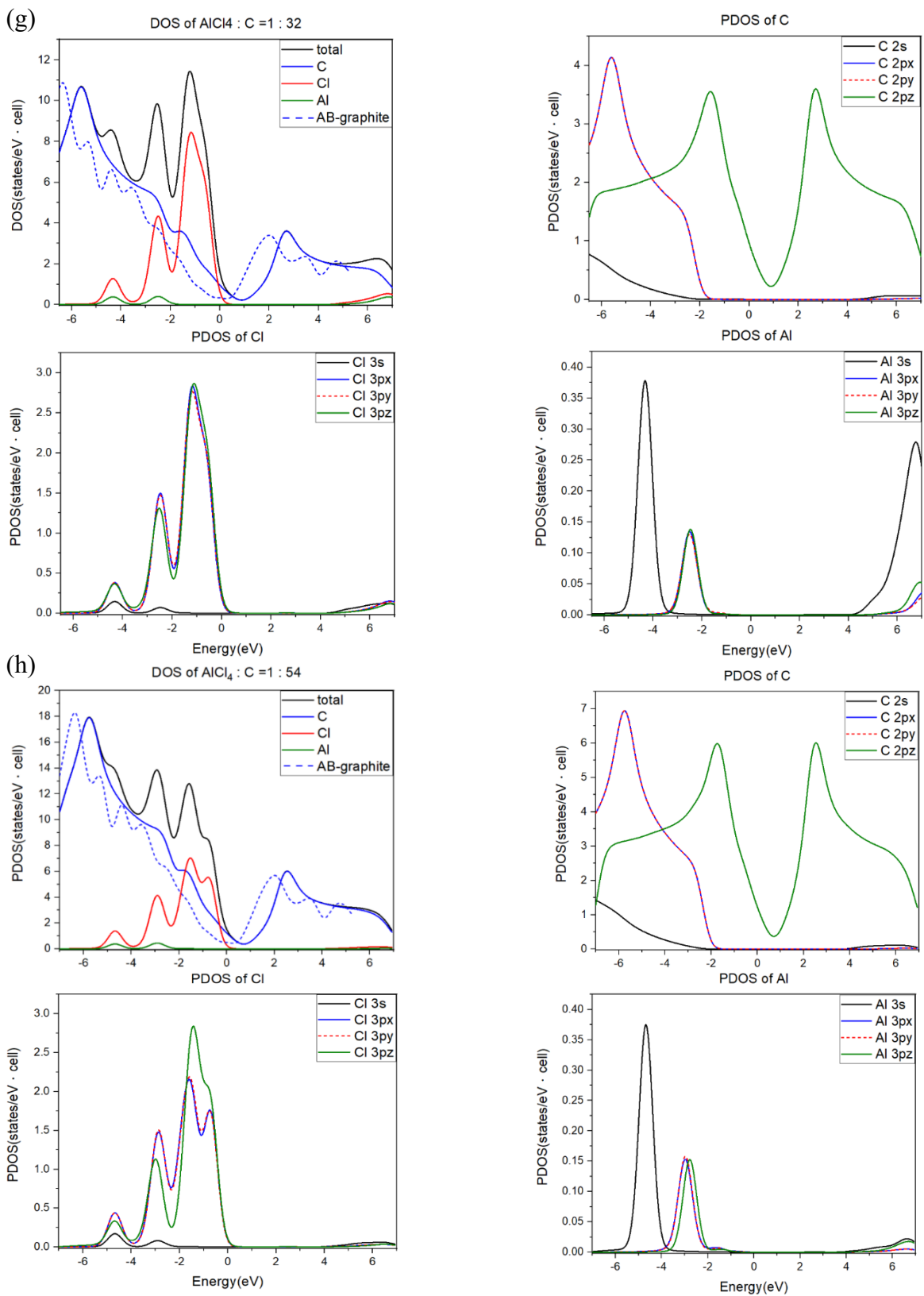

Figure 4: The atom- & orbital-projected density of states for $AlCl_4^-/AlCl_4$-intercalation graphite compounds under different molecular concentrations: (a)/(e) 1:18, (b)/(f) 1:24, (c)/(g) 1:32 and (d)/(h) 1: 54.